\documentclass{article}
\usepackage{dcase2022,amsmath,graphicx,url,times,booktabs, tabularx, enumitem}

\title{Impact of temporal resolution on convolutional recurrent networks\\ for audio tagging and sound event detection}

\name{Wim Boes, Hugo Van hamme\thanks{This work was supported by a PhD Fellowship of Research Foundation Flanders (FWO-Vlaanderen) and the Flemish Government under ``Onderzoeksprogramma AI Vlaanderen''.}}
\address{
ESAT, KU Leuven, Belgium}
\begin{document}

\ninept
\maketitle

\begin{sloppy}

\begin{abstract}
Many state-of-the-art systems for audio tagging and sound event detection employ convolutional recurrent neural architectures. Typically, they are trained in a mean teacher setting to deal with the heterogeneous annotation of the available data. 

In this work, we present a thorough analysis of how changing the temporal resolution of these convolutional recurrent neural networks --- which can be done by simply adapting their pooling operations --- impacts their performance. By using a variety of evaluation metrics, we investigate the effects of adapting this design parameter under several sound recognition scenarios involving different needs in terms of temporal localization.
\end{abstract}


\section{Introduction}
\label{sec:intro}

Recently, the popularity of research into audio-related tasks has surged because of, among other reasons, multiple versions of the Detection and Classification of Acoustic Scenes and Events (DCASE) challenge~\cite{DCASE2016, DCASE2017proceedings, DCASE2018proceedings, DCASE2019proceedings, DCASE2020proceedings, DCASE2021proceedings}. The latest editions encompassed six categories, each dealing with a distinct sound recognition problem. Subtask number 4~\cite{Turpault} consistently covered sound event detection --- joint classification and temporal localization of auditory events --- in domestic environments. Submitted systems were ranked utilizing measures which represent scenarios involving variable requirements with regard to the estimation of time boundaries.

The baseline~\cite{Turpault} supplied for the fourth problem of the DCASE 2021 and 2022 challenges, a convolutional recurrent neural network trained using the mean teacher principle~\cite{meanteacher}, was based on the second-ranking system~\cite{dcase2019winner} of task 4 of DCASE 2019. Its output temporal resolution was predefined and fixed.

Our submission~\cite{boesreport} for the fourth subtask of the DCASE 2021 challenge~\cite{Turpault} demonstrated that simply adapting the amount of pooling in this network, which is directly linked to its temporal resolution, can significantly impact its performance.

In this work, we provide a more thorough and complete analysis of this interesting finding. We substantially supplement the discussion in multiple ways, as outlined in the two paragraphs below. 

Firstly, we examine more than the specific evaluation scenarios prescribed in task 4 of the DCASE 2021 and 2022 challenges. Particularly, we also investigate what happens when the temporal resolutions of convolutional recurrent networks are adapted in the context of audio tagging. In this situation, the goal of the models is to perform clip-level auditory event classification. Unlike for sound event detection, temporal localization is not required in this case.

Secondly, for sound event detection, we inspect what happens using multiple types of measures. More specifically, in addition to the intersection-based scores employed in subtask 4 of the DCASE 2021 and 2022 challenges, we also utilize segment-based and event-based metrics, which are commonly used in this subdomain of machine learning as well. This is further elaborated upon in Section~\ref{sect:setup}.

To this end, the following approach is taken: We start with a suitable baseline, which forms the basis for many state-of-the-art architectures for both audio tagging and sound event detection, such as \cite{sota1}, \cite{sota2} and \cite{sota3}. We change the temporal resolution of this model by adapting its pooling operations. We then analyze the results obtained by this modified system in a variety of experimental configurations, representing different evaluation scenarios.

In Section~\ref{sect:baseline}, we expand upon the baseline system. Afterwards, in Section~\ref{sect:tempres}, we describe how the pooling operations of the considered architecture can be adapted to modify the temporal resolution of the model. Then, in Section~\ref{sect:setup}, we elaborate upon the experimental setup. Next, in Section~\ref{sect:results}, we analyze the results of the performed experiments, and finally, we draw a conclusion in Section~\ref{sect:conclusion}.

\section{Baseline}
\label{sect:baseline}

In this section, the baseline system, which is nearly the same as in \cite{interspeech}, is elaborated upon. It is a slightly adapted version of the baseline supplied for task 4 of the DCASE 2021 and 2022 challenges~\cite{Turpault}.

\subsection{Architecture}

A schematic visualization of the baseline model is given in Figure~\ref{fig:baseline}.

The input to the baseline is a spectral map of an audio recording. The amount of frequency bins is predefined and set to 128.

The first component of the architecture is a convolutional neural network (CNN) made up of seven blocks. Each of those consists of the following five layers: a convolutional layer, a batch normalization layer~\cite{batchnorm}, a ReLU activation layer, a dropout layer~\cite{dropout} with a dropout rate of 33\%, and lastly, an average pooling layer. 

All convolutional layers use a square kernel of size 3 and utilize a stride of 1. For the first three blocks, the number of output channels of the convolutional operations is equal to 16, 32 and 64 respectively. For the last four blocks, this number is equal to 128.

The hyperparameters of the pooling operations are given per model block in Table~\ref{tab:poolhyper}. The first and second numbers of each tuple are connected to the time and frequency axes respectively.

After the last block, the frequency-related dimension of the spectral (auditory) input map has been brought down to one and the corresponding axis can therefore be squeezed, i.e., removed. 

Afterwards, a bidirectional gated recurrent (BiGRU) unit, consisting of two layers with hidden sizes equal to 128, is used to model potential temporal relationships in the auditory data. 

\begin{figure}[!ht]
\centering
\includegraphics[width=6.70800cm]{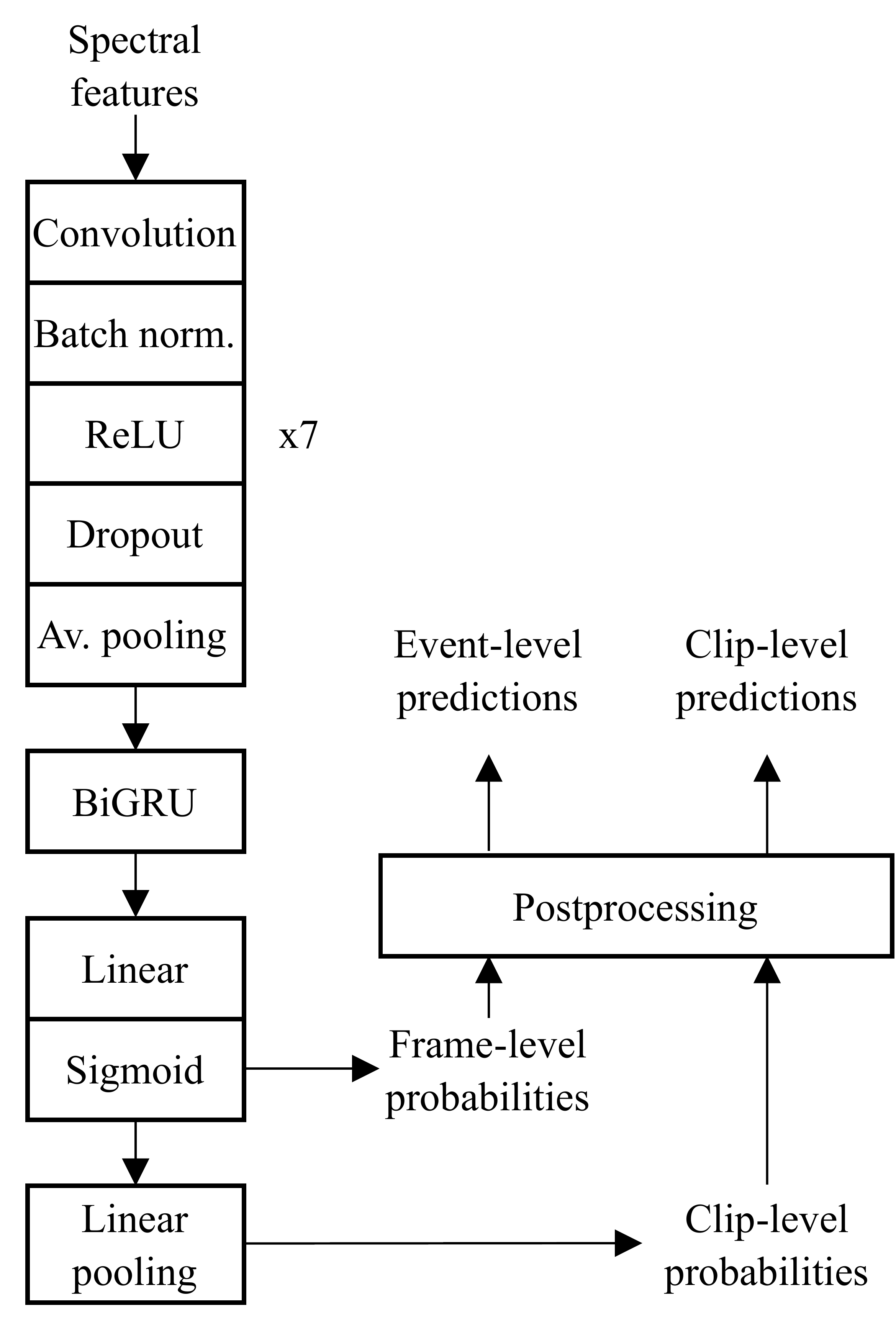}
\caption{Schematic representation of the baseline}
\label{fig:baseline}
\end{figure}

The output of this recurrent layer is followed by a linear projection and application of the sigmoid function to obtain multi-label frame-level probabilities. These values indicate per temporal frame which sound categories are active, and can be postprocessed to obtain event-level predictions of sounds, which are relevant for sound event detection, as explained in the next subsection. 

Finally, the frame-level sound probabilities are aggregated to get clip-level auditory event probabilities, which are relevant for the task of audio tagging, by performing linear (softmax) pooling~\cite{poolcomparison}.

\subsection{Postprocessing}

The clip- and frame-level probabilities are converted into binary values by enforcing a fixed threshold. The frame-level decisions are also passed through a smoothing median filter with a window of about 500 ms. Preliminary experiments showed that more complicated postprocessing (e.g., class-wise optimization on validation data) is unnecessary as it provides relatively insignificant benefits. 

The binary clip-level decisions can readily be used to perform audio tagging. On the contrary, to produce outputs suitable for sound event detection, an extra step has to be taken: The frame-level decisions are converted into event-level predictions by performing a merging operation with a maximum gap tolerance of 200 ms.

\subsection{Mean teacher training}

As expanded upon in Section~\ref{sect:setup}, the training data employed in this research project is heterogeneously annotated: Some of the samples include clip-level or weak labels, some come with event-level or strong labels, and the rest is unlabeled. To deal with this difficulty, the mean teacher training principle~\cite{meanteacher} is applied. 

\begin{table}[!ht]
\caption{Kernel sizes and strides of pooling layers in baseline CNN}
\label{tab:poolhyper}
\centering
\begin{tabular}{@{}lc@{}}
\toprule
\textbf{Block} & \textbf{Kernel size $=$ stride} \\
\midrule
0-1 & (2, 2) \\
2-3-4-5-6 & (1, 2) \\
\bottomrule
\end{tabular}
\end{table} 

In the mean teacher training framework, two models called the student and the teacher are utilized. They share the same architecture, but their parameters are updated completely differently.

The student system is trained in a regular fashion: A differentiable loss function is optimized by an optimization algorithm such as Adam~\cite{adam}. This is not the case for the teacher counterpart: The weights of this model are computed as the exponential moving average of the student parameters with a multiplicative decay factor of 0.999 per training iteration, also explaining the name of the method.

The loss used to train the student consists of four terms: The first two are clip-level and frame-level binary cross entropy functions, which are only calculated for the weakly and strongly labeled data samples respectively. The remaining two components are mean-squared error consistency costs between the clip-level and frame-level output probabilities of the student and teacher models, which can be computed for all examples, including the unlabeled ones. The classification and consistency terms are summed with weights equal to 1 and 2 respectively to obtain the final objective. 

\section{Adaptation of temporal resolution}
\label{sect:tempres}

The goal of this project is to investigate how modifying the output temporal resolution of the considered convolutional recurrent neural network affects its sound recognition performance under different circumstances. In this section, we describe which parts of this model were changed to achieve these specific adaptations.

In the baseline model outlined in Section~\ref{sect:baseline}, the pooling layers in the first and second blocks of the CNN halve input feature maps along the temporal dimension, while the others do not change anything in this regard. This results in a total pooling factor of 4.

As explained in more detail in Section~\ref{sect:setup}, most of the audio recordings used in this project have a duration of 10 seconds. They are fed to the baseline convolutional recurrent neural network as spectral feature maps consisting of 608 time frames. This model applies a temporal reduction factor of 4, and hence, the frame-level probabilities contain 152 values per sample. This comes down to a temporal resolution of about one class prediction vector per 65 ms. 

We create adaptations of the baseline convolutional recurrent network in the following way: Instead of only allowing the first two pooling layers to apply a reduction in the number of temporal frames, we create model versions of which the first $x$ pooling layers perform this halving operation, with $x$ ranging from 1 up to 6. This results in systems with time reduction factors of 2, 4, 8, 16 and 32 respectively, which are equivalent to temporal resolutions of about one frame-level prediction bin per 32.5, 65, 130, 260 and 520 ms.

Important to note is that changing the amount of pooling in the considered convolutional recurrent neural network does not change the number of trainable parameters, and thus, its modeling capacity.

\section{Experimental setup}
\label{sect:setup}

In this section, we provide full details on the setup used during the experiments, of which the results are analyzed in Section~\ref{sect:results}.

\subsection{Data}
\label{sect:data}

The data set for problem 4 of DCASE 2021~\cite{Turpault} is a multi-label collection consisting of audio recordings with a maximum length of 10 seconds. There are 10 possible sound event categories, related to domestic environments, which are not mutually exclusive. 

The partition used to train models consists of three subsets:
\begin{itemize}[noitemsep, topsep=0pt]
    \item 1578 real samples with weak (clip-level) annotation
    \item 10000 synthetic samples with strong (event-level) annotation
    \item 14412 real samples without annotation
\end{itemize}

For evaluation purposes, two partitions were available: the so-called validation and public evaluation sets, holding 1168 and 692 real recordings respectively. Both subsets include strong or event-level labels of the active environmental auditory events.

\subsection{Preprocessing}

To get spectral maps of the available audio recordings, which are used as input features for the considered convolutional recurrent nets as explained in Section~\ref{sect:baseline}, we resampled all clips to 22050 Hz and performed peak amplitude normalization. Then, log mel spectrograms with 128 frequency bins were extracted using a Hamming window with a size of 2048 samples and a hop length of 363 samples. Lastly, per-frequency bin standardization was carried out.

As mentioned before, for a 10-second audio clip, these steps resulted in a spectral feature map consisting of 608 temporal frames.

\subsection{Data augmentation}

In order to avoid the risk of overfitting, we employed data augmentation during the training of the considered models. In particular, we used mixup~\cite{mixup}, which comes down to creating extra learning examples (and associated labels) by linearly interpolating the original samples. We employed this method with a probability of 50\% of applying it. The mixing ratios used in this algorithm were randomly sampled from a beta distribution with shape parameters set to 0.2.

Unlike for our related submission~\cite{boesreport} for task 4 of the DCASE 2021 challenge~\cite{Turpault}, we did not employ time and frequency masking~\cite{specaugment} in this project. This choice was based on two observations made during initial experiments: Firstly, the best hyperparameters for these techniques seemed to depend on the situation, e.g., used metric. Secondly, these methods only led to minimal improvements. As the focus of this work is on analysis rather than trying to get optimized performance, they were excluded for the sake of clarity.

\subsection{Training and evaluation}

All models were trained and evaluated using PyTorch~\cite{pytorch}.

\subsubsection{Training}

All models were trained for 200 epochs. Per epoch, 250 batches of 48 samples were given to the networks. Each batch contained 12 weakly labeled, 12 strongly labeled and 24 unlabeled examples.

Adam~\cite{adam} was employed to train the weights of the student models. Learning rates were ramped up exponentially from 0 to 0.001 for the first 12500 optimization step. Thereafter, they decayed multiplicatively at a rate of 0.99995 per training iteration.

\subsubsection{Evaluation}

For audio tagging, we used the (micro-averaged) clip-based F1 measure~\cite{metrics}, which was also utilized in task 4 of the DCASE 2017 challenge~\cite{DCASE2017challenge}. This score was computed for a single operating point, namely, the situation in which probabilities were converted into binary decisions by enforcing a 50\% threshold.

For sound event detection, we employed multiple types of metrics. To start with, we utilized the intersection-based measures used for ranking in task 4 of the DCASE 2021 and 2022 challenges~\cite{Turpault}. More specifically, we used two polyphonic sound event detection scores~\cite{psds} representing distinct evaluation scenarios, denoted as PSDS 1 and 2. The former imposes strict requirements on the temporal localization accuracy, the latter is more lenient in this regard. 

The hyperparameters utilized for calculating these PSDS measures are summarized in Table~\ref{tab:psdshyper}. These scores were computed using 50 operating points, in which thresholds linearly distributed from 0.01 to 0.99 were used to convert probabilities into binary decisions. 

Details on the procedure for calculating PSDS scores and a discussion on the hyperparameters can be found in~\cite{psds}. 

\begin{table}[!ht]
\caption{PSDS hyperparameters}
\label{tab:psdshyper}
\centering
\begin{tabular}{@{}lcc@{}}
\toprule
\textbf{Hyperparameter} & \textbf{PSDS 1} & \textbf{PSDS 2} \\
\midrule 
Detection tolerance criterion & 0.7 & 0.1 \\
Ground truth intersection criterion & 0.7 & 0.1 \\
Cross-trigger tolerance criterion & N/A & 0.3 \\
Cost of class instability & 1 & 1 \\
Cost of cross-triggers & 0 & 0.5 \\
Maximum false positive rate & 100 & 100 \\ 
\bottomrule
\end{tabular}
\end{table}

Furthermore, we used the segment-based (micro-averaged) F1 score based on chunks of 1 s~\cite{metrics} to gauge the performance of the considered models. This metric was also employed in task 4 of the DCASE 2017 challenge~\cite{DCASE2017challenge}. Because of the long segment length, this measure quantifies systems in terms of their ability to perform more coarse-grained sound event detection, not unlike PSDS 2.

Lastly, we also employed the (macro-averaged) event-based F1 score with tolerances of 200 ms for onsets and 20\% of the audio event lengths (up to a max of 200 ms) for offsets~\cite{metrics}. This measure was used in task 4 of the DCASE 2018, 2019 and 2020 challenges~\cite{Turpault, Serizel}. This metric quantifies models in terms of their ability to perform fine-grained sound event detection, similar to PSDS 1.

These segment-based and event-based scores were computed for a single operating point, in which a threshold of 0.5 was enforced to convert frame-level probabilities into binary decisions. 

All measures were computed using the clip-level or frame-level probabilities of the student models after the last training epoch.

\section{Experimental results}
\label{sect:results}

In this section, we analyze the results obtained by the convolutional recurrent neural systems with varying output resolutions, as explained previously. We report the metrics for audio tagging and sound event detection elaborated upon in Section~\ref{sect:setup} after applying the following method to significantly diminish the variability of the results: For each experimental configuration, we train 20 models with independent initializations and average the scores they achieve on the public evaluation partition of the employed data set.

\begin{figure}[!ht]
\centering
\includegraphics[width=8.5cm]{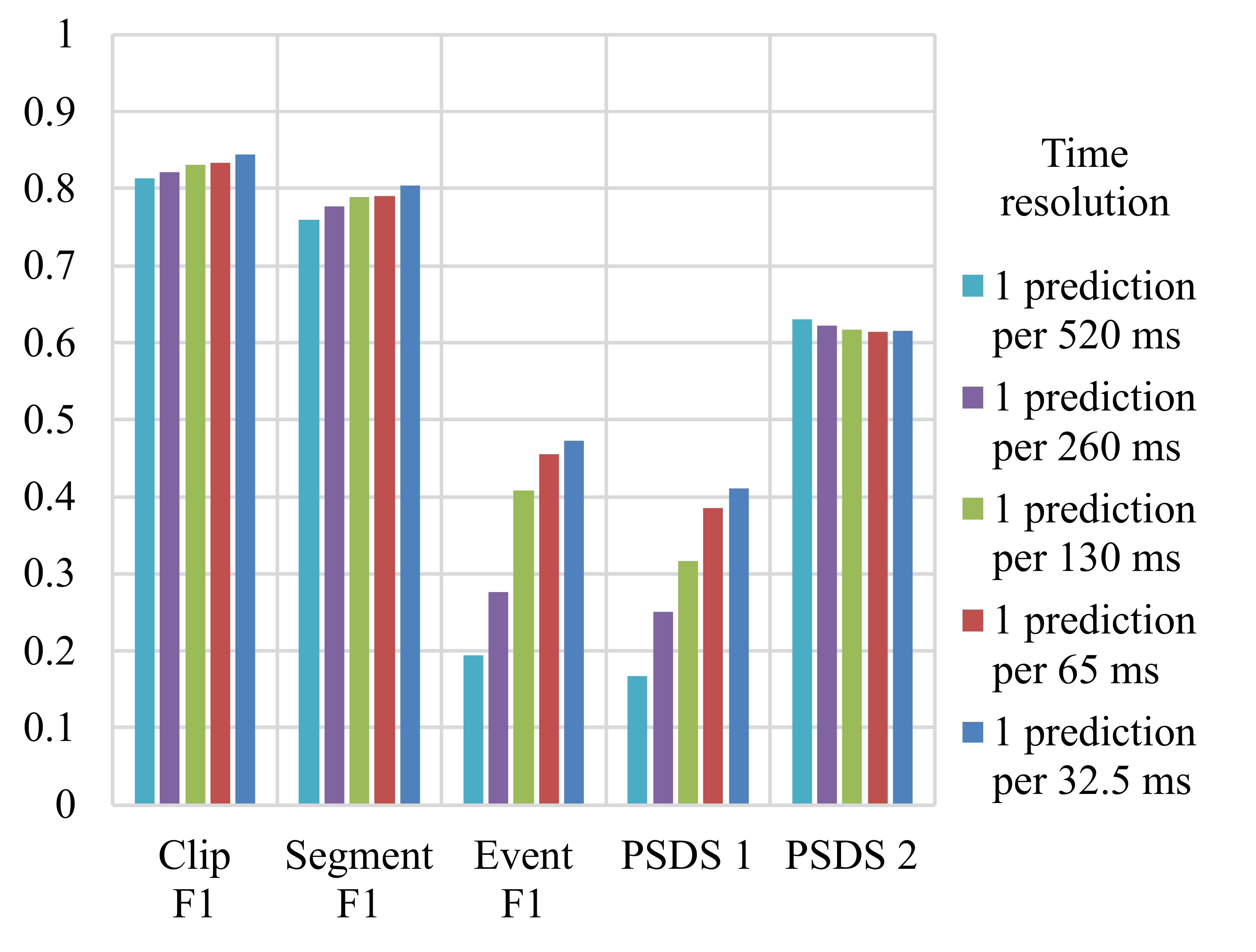}
\caption{Results on public evaluation subset}
\label{fig:test}
\end{figure}

The scores achieved on the public evaluation set are summarized in Figure~\ref{fig:test}. Precise numerical values are not essential to the interpretation below and are not disclosed due to a lack of space. 

Only one of the metrics appears to display a (very slight) negative correlation with the temporal output resolution of the considered models, namely, PSDS 2. Intuitively, this behavior is not unexpected as this intersection-based score was designed to gauge relatively coarse-grained sound event detection performance. 

All other measures correlate positively with respect to the time resolution. For the event-based F1 score and PSDS 1, the scores demanding precise estimations of the boundaries of sounds, this seems logical: Naturally, if the length associated with an output prediction bin is too long, this becomes more challenging or even impossible. 

However, this positive relationship also appears to hold for the clip-based and segment-based F1 metrics, which inherently require much less accurate temporal differentiation --- in the former case, this is not even needed at all. This counterintuitive result is all the more unanticipated given the trend of PSDS 2: Apparently, there is a certain discrepancy in the reaction of the examined types of measures for audio tagging and coarse-grained sound event detection to the temporal output resolution of the considered models.

The sensitivity of the performance of convolutional recurrent neural networks to their output resolutions depends on the criterion. In particular, scores designed for more fine-grained sound event detection show a high responsiveness in this regard: PSDS 1 and the event-based F1 measure drop sharply as the temporal resolution of the models at hand decreases. On the contrary, the clip- and segment-based F1 metrics as well as PSDS 2, used to gauge performance with regard to audio tagging and coarse-grained sound event detection, remain more stable as this hyperparameter varies.

\begin{figure}[!ht]
\centering
\includegraphics[width=8.5cm]{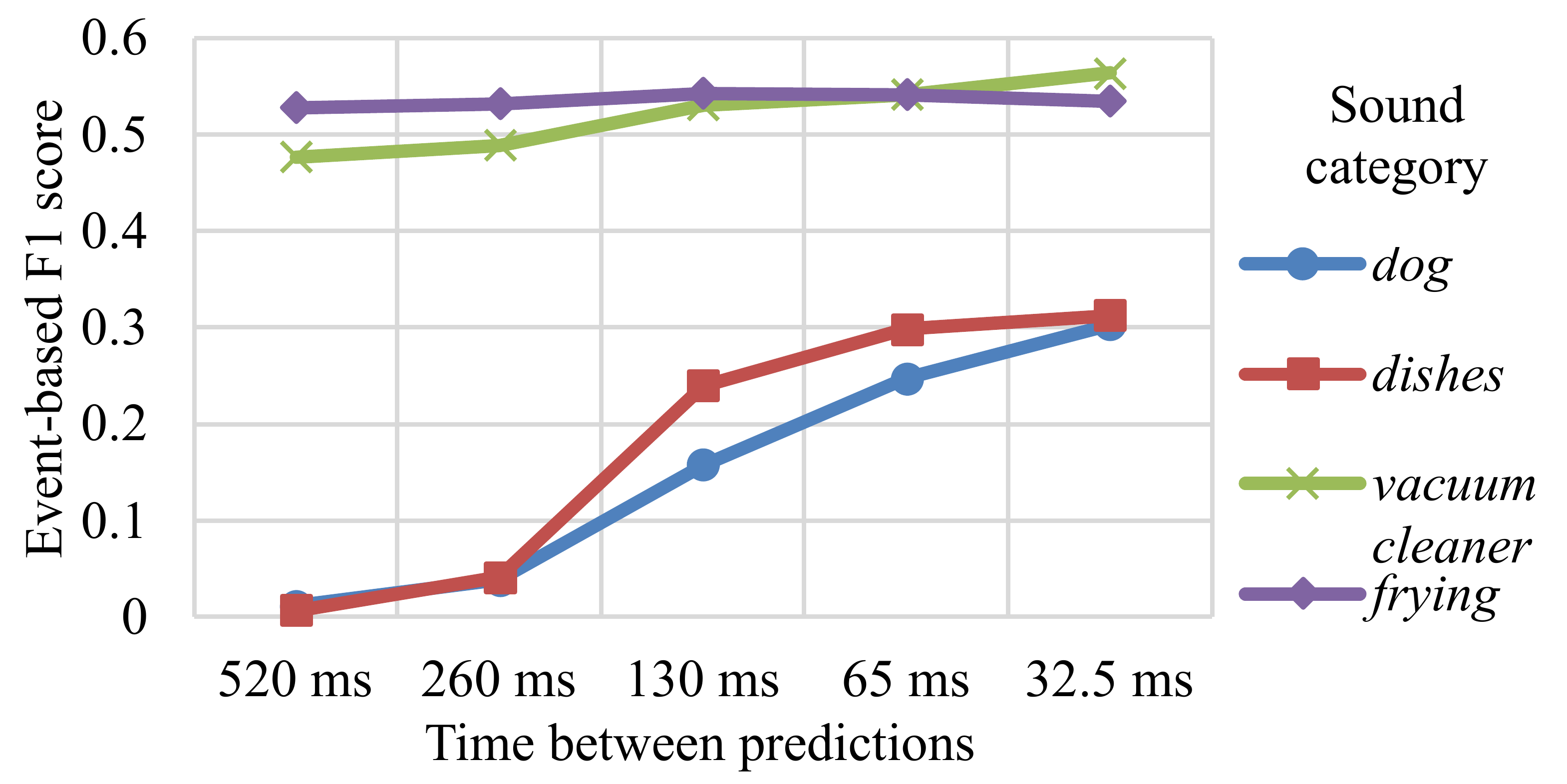}
\caption{Subset of class-wise event-based F1 scores}
\label{fig:class}
\end{figure}

It is also possible to study how the performance and sensitivity of the models differs across sound categories by inspecting class-wise scores. For F1-based measures, these can easily be obtained. For the PSDS metrics, they can also be computed, but unlike for their aggregated counterparts, the cost of class instability (see Section~\ref{sect:setup}) must be disregarded. Exact numerical results are not crucial to the following analysis and are not included due to a lack of space.

In nigh all instances, the class-wise scores correlate to the temporal resolution in the same direction (positively/negatively) as their global versions. Nevertheless, the absolute performance and responsiveness strikingly vary across metrics and sound categories.

For all of the measures, there is a substantial link between the scores and the examined kind of sound. As an example, the best-performing category (\emph{speech}) consistently outperforms the worst (\emph{running water}) by a large margin. This behavior can be explained by multiple factors, e.g., qualitative differences between types of audio events and imbalance within the data used for training.

As is the case for the aggregated scores, the class-wise sensitivities are only strongly pronounced for fine-grained sound event detection measures. In these cases, they also greatly depend on the considered audio category. This is exemplified for the event-based F1 metric in Figure~\ref{fig:class}. For events which are short and localized (e.g., \emph{dog}, \emph{dishes}), reducing the temporal resolution leads to severe deterioration in terms of performance. For more long-lived sounds (e.g., \emph{vacuum cleaner}), this dependency is much less outspoken. In the most extreme case of \emph{frying}, there is even no correlation at all. 

\section{Conclusion}
\label{sect:conclusion}

Numerous state-of-the-art sound recognition models are based on convolutional recurrent neural architectures. They are usually optimized in a mean teacher framework to deal with the heterogeneous labeling of the supplied data samples. In this work, we provided deeper insight into how changing the temporal resolution of such nets impacts their performance. We analyzed the effect of modifying pooling operations on a multitude of metrics, designed for distinct audio tagging and sound event detection scenarios.

The experiments showed that the performance of the considered models is highly susceptible to changes in terms of their pooling operations. More specifically, metrics designed for fine-grained sound event detection showed a strong, positive relation with respect to the temporal output resolutions of the systems at hand. For sound recognition measures involving less focus on temporal differentiation (among others, F1 score for audio tagging), the direction of the correlation was mixed and contingent on the scoring method.

The responsiveness appeared to depend on the considered evaluation scenario. Generally, measures devised for fine-grained sound event detection displayed a much higher reactivity than metrics built for audio tagging and coarse-grained sound event detection.

A class-wise study showed that for the latter, the sensitivity to temporal resolution was low across all sound types. Conversely, for fine-grained sound event detection scores, the reactivity was decidedly stronger for shorter than for longer-lasting events.

\bibliographystyle{IEEEtran}
\bibliography{refs}

%
%
%
%
%
%
%
%
%

\end{sloppy}
\end{document}